\documentclass[twocolumn]{aastex631}
\usepackage{xcolor}

\begin{document}

\title{A Metallicity Catalog of Very Metal-poor Main-sequence Turn-off and Red Giant Stars from LAMOST DR10}

\author{Xiangyu Li}
\affiliation{National Astronomical Observatories, Chinese Academy of Sciences, Beijing 100101, China}
\affiliation{School of Astronomy and Space Science, University of Chinese Academy of Sciences, Beijing 100049, China}

\author[0009-0008-2988-2680]{Huiling Chen}
\affiliation{Department of Astronomy, School of Physics, Peking University, Beijing 100871, China}
\affiliation{Kavli Institute for Astronomy and Astrophysics, Peking University, Beijing 100871, China}

\author{Yang Huang}
\affiliation{School of Astronomy and Space Science, University of Chinese Academy of Sciences, Beijing 100049, China}
\affiliation{National Astronomical Observatories, Chinese Academy of Sciences, Beijing 100101, China}

\author{Huawei Zhang}
\affiliation{Department of Astronomy, School of Physics, Peking University, Beijing 100871, China}
\affiliation{Kavli Institute for Astronomy and Astrophysics, Peking University, Beijing 100871, China}

\author{Timothy C. Beers}
\affiliation{Department of Physics and Astronomy and JINA Center for the Evolution of the Elements (JINA-CEE), University of Notre Dame, Notre Dame, IN 46556, USA}

\author{Linxuan Zhu}
\affiliation{Department of Astronomy, School of Physics, Peking University, Beijing 100871, China}
\affiliation{Kavli Institute for Astronomy and Astrophysics, Peking University, Beijing 100871, China}

\author{Jifeng Liu}
\affiliation{National Astronomical Observatories, Chinese Academy of Sciences, Beijing 100101, China}
\affiliation{School of Astronomy and Space Science, University of Chinese Academy of Sciences, Beijing 100049, China}

\correspondingauthor{Yang Huang (huangyang@ucas.ac.cn); Huawei Zhang (zhanghw@pku.edu.cn)}

\begin{abstract}

We present a catalog of 8,440 candidate very metal-poor (VMP; [Fe/H] $\leq -2.0$) main-sequence turn-off (MSTO) and red giant stars in the Milky Way, identified from low-resolution spectra in LAMOST DR10. More than 7,000 of these candidates are brighter than $G \sim 16$, making them excellent targets for high-resolution spectroscopic follow-up with 4–10 meter-class telescopes.
Unlike most previous studies, we employed an empirical calibration to estimate metallicities from the equivalent widths (EWs) of the Calcium Triplet (CaT) lines, taking advantage of the high signal-to-noise ratio (SNR) in the red arm of LAMOST spectra. We further refined this calibration to improve its reliability for more distant stars. This method enables robust identification of VMP candidates with metallicities as low as [Fe/H] = $-4.0$ among both MSTO and red giant stars.
Comparisons with metal-poor samples from other spectroscopic surveys and high-resolution follow-up observations confirm the accuracy of our estimates, showing a typical median offset of $\sim$0.1 dex and a standard deviation of $\sim$0.2 dex.
\end{abstract}


\section{Introduction \label{sec:1}}

Very metal-poor (VMP; [Fe/H]~$\leq –2.0$) stars serve as fossil records of the early chemical enrichment and dynamical assembly of the Milky Way. Their chemical compositions offer valuable clues about the star-formation history of the early Universe, as well as the nature of the Big Bang, the initial mass function, the nucleosynthetic yields of the first generations of stars, and the role of early supernovae in shaping the interstellar medium \citep{Beers, IMF, TheFirstStars}. In addition, the dynamical properties of VMP stars provide key insights into the formation of the Galactic halo and the major merger events that occurred during the Galaxy’s early assembly history (e.g., \citealt{Shank2022, Shank2023, Zepeda2023, Cabrera2024}, and references therein).

Due to their rarity and scientific significance, the search for VMP stars has long been a challenging yet central pursuit in Galactic Archaeology. Over the past decade, large-scale spectroscopic and photometric surveys have enabled unprecedented opportunities for their discovery. Millions of VMP candidates have been identified through data mining of spectroscopic surveys such as RAVE \citep{RAVE}, LAMOST \citep{LAMOST, LAMOST2}, SEGUE \citep{SEGUE, Rockosi2022}, APOGEE \citep{APOGEE}, GALAH \citep{GALAH}, and Gaia \citep{Gaia}, objective-prism surveys such as the HK survey \citep{Beers1985, Beers1992} and the Hamburg/ESO survey \citep{Christlieb2008}, and photometric surveys like the Pristine survey \citep{Pristine}, the SkyMapper survey \citep{SkyMapper, HuangSMSS1, Huang1}, J-PLUS \citep{JPLUS, Huang3}, S-PLUS \citep[][ Y. Huang et al. in prep.]{Mendes2019}, and SAGES \citep{SAGES, Huang2}. However, many of these samples are based on survey pipelines or machine-learning methods, which generally struggle to identify stars below [Fe/H]~$\sim –2.5$ due to the intrinsic weakness of metallic lines at such low abundances. In the case of the photometric surveys, the influence of strong molecular carbon bands can lead to incorrect (high) metallicity estimates (see, e.g., \citealt{Hong2024}).

Although high-resolution spectroscopic follow-up provides precise measurements, the number of VMP stars confirmed through such observations remains relatively small. A major limitation lies in the low success rate of candidate selection from low-resolution or photometric data—many turn out not to be truly metal-poor. This leads to a significant waste of valuable high-resolution resources and hinders the construction of reliable VMP samples. These challenges underscore the urgent need for a large and reliable sample of VMP candidates that can efficiently guide follow-up efforts toward the most promising targets.

\begin{figure}
\centering
\plotone{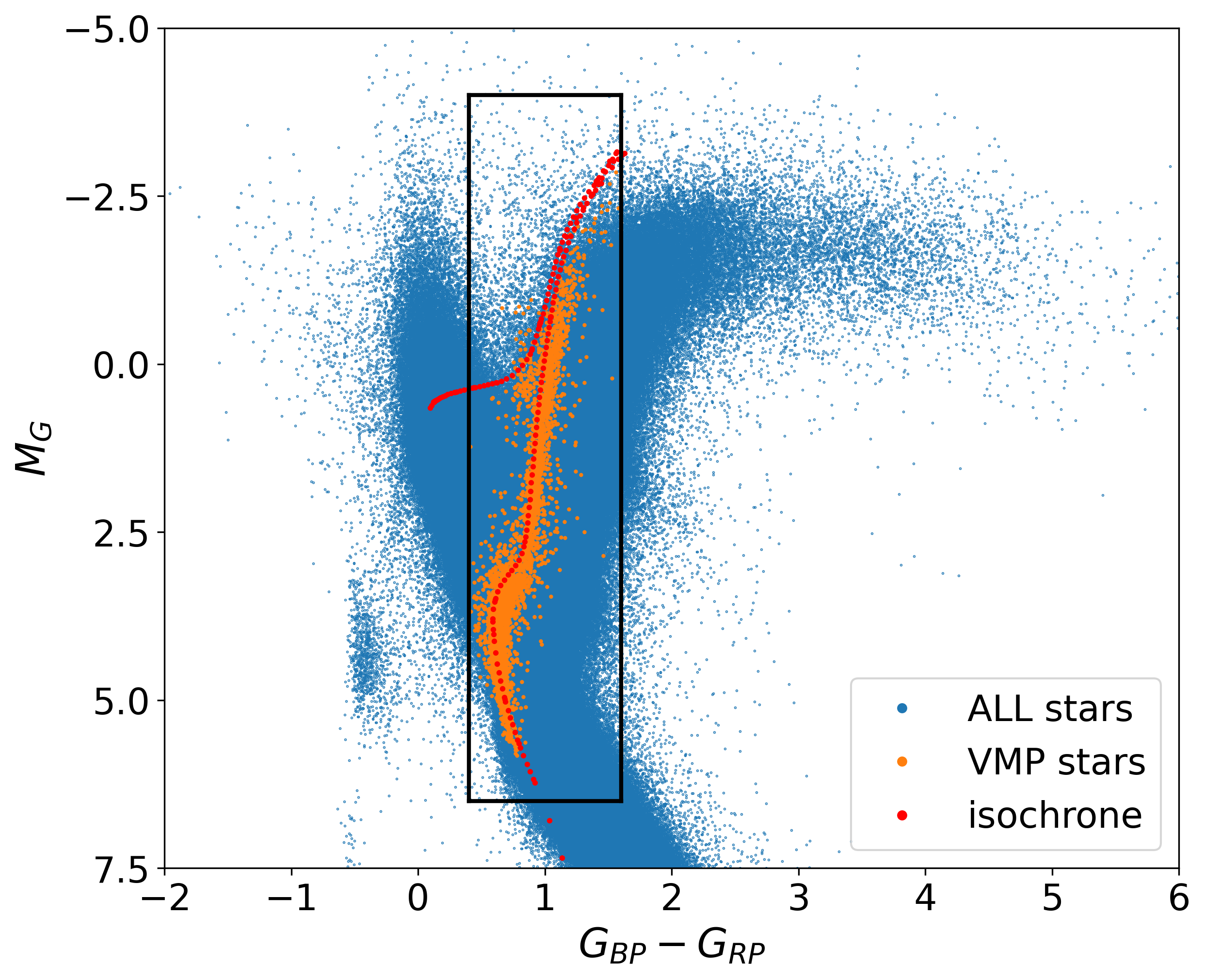}
\caption{Color-magnitude diagram (CMD) of stars in LAMOST DR10. The black rectangle indicates our empirical cut for MSTO and red giant stars, defined by $0.4 < G_{BP}-G_{RP} < 1.6$ and $-4.0 < M_G < 6.5$. The blue data points represent the full stellar sample from LAMOST DR10, while the orange data points represent the VMP stars identified by Method 1. These stars exhibit a good agreement with the PARSEC isochrone for [Fe/H] $=-$2.0 and an age of 10 Gyr, shown as the red-dotted line.}
\label{fig:HR}
\end{figure}

LAMOST DR10 has released over 12 million low-resolution stellar spectra across the Milky Way, offering a rich dataset for stellar parameter and chemical-abundance analysis. With its wide wavelength coverage (3700--9100\,\AA) and sufficient spectral resolution ($R \sim 1800$), LAMOST enables the determination of multiple elemental abundances and atmospheric parameters \citep{LAMOST2}. Most existing pipelines, such as those described in \citet{AFGK} and \citet{Haining}, primarily focus on the blue arm of the spectrum, where metal absorption lines are abundant. However, for VMP stars, the blue spectral region often suffers from low signal-to-noise ratios (SNR), making metallicity measurements challenging due to the intrinsic weakness of metal lines, and/or poor determination of the continuum. In contrast, the red arm of LAMOST spectra typically exhibits significantly higher SNR, which is crucial for detecting and characterizing distant and faint VMP stars. Fully leveraging this high-SNR red spectral range allows for the identification of a larger number of reliable VMP candidates than previously possible, and improves the precision of metallicity estimates for already known ones.

Since the red arm of LAMOST spectra contains fewer spectral features than the blue arm, we abandoned the synthetic spectral fitting methods commonly adopted by LAMOST pipelines and employ an alternative approach. We focus on the Calcium Triplet (CaT) lines ($\lambda = 8500$, 8544, 8664 \AA), which are among the most prominent features in the near-infrared region of LAMOST spectra. \citet{Carrera} demonstrated that the CaT lines serve as reliable metallicity indicators for VMP red giant stars, and provided a simple empirical calibration capable of measuring metallicities down to [Fe/H] $\sim -$4.0 using their equivalent widths (EWs). \citet{Gaiafeh} further applied this method to \textit{Gaia} RVS spectra, showing excellent agreement with high-resolution spectroscopic measurements, confirming its robustness.

\begin{table}
\caption{Coefficients for the Calibrations used in this Work }
\begin{tabular}{crr}
\hline
\textbf{Coefficients} & \textbf{Method 1} & \textbf{Method 2} \\
\hline
\textit{a} & $-$3.45$\pm$0.04  & $-$3.68$\pm$0.15 \\
\textit{b} &    0.16$\pm$0.01  & $-$0.72$\pm$0.12 \\
\textit{c} &    0.41$\pm$0.004 &    0.49$\pm$0.01 \\
\textit{d} & $-$0.53$\pm$0.11  &    0.35$\pm$0.02 \\
\textit{e} &   0.019$\pm$0.002 & \dots \\
\hline
\end{tabular}
\vskip 0.25cm
Coefficients in Method 1 are taken from \citet{Carrera}.

\label{tab:1}
\end{table}

\begin{figure*}
\centering
\plotone{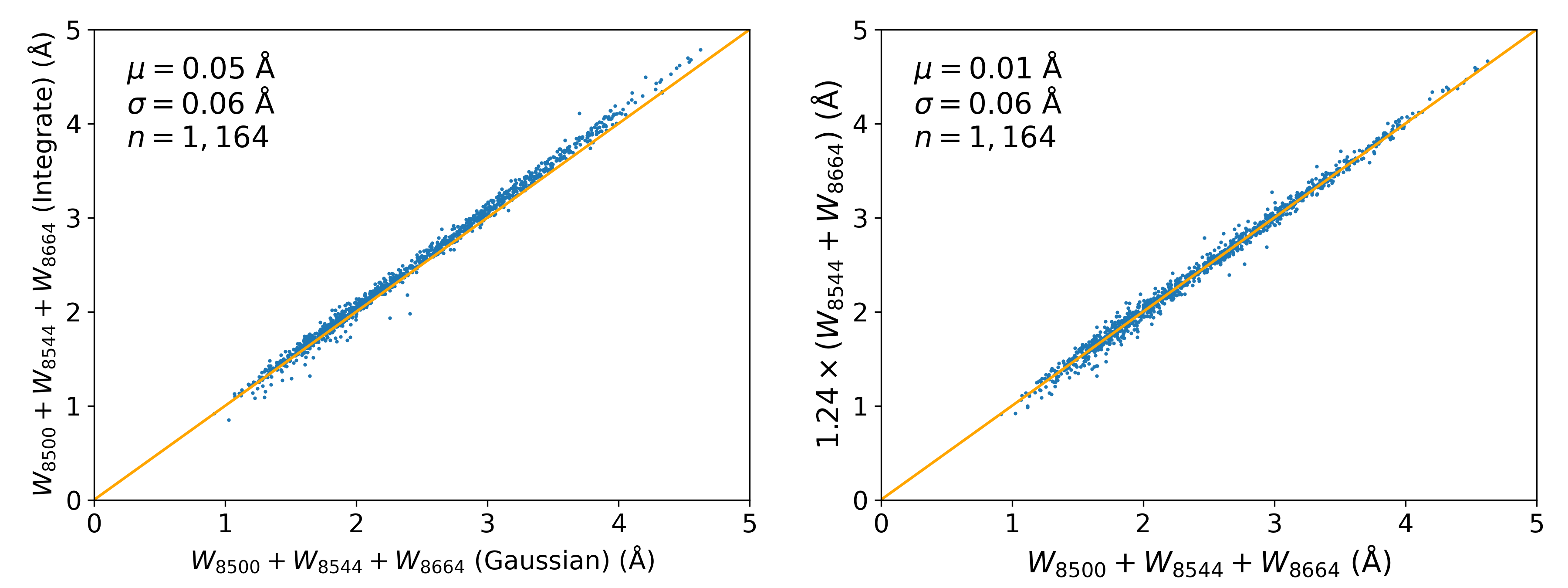}
\caption{Left: Comparison of the sum of the EWs of three CaT lines, $W_{8500}+W_{8544}+W_{8664}$, for the high-SNR VMP sample using the Gaussian fitting method and direct integration method. Right: Comparison of two definitions of $\Sigma$Ca: the sum of the three lines $W_{8500}+W_{8544}+W_{8664}$, and 1.24 times the sum of the last two lines $1.24 \times (W_{8544}+W_{8664})$. The number of samples (n), median offset ($\mu$), and standard deviation ($\sigma$) are shown at the top. The orange line represents the 1:1 line.}
\label{fig:ew}
\end{figure*}

Building on these studies, we adopt the empirical calibration of \citet{Carrera} to estimate metallicities for red giant stars in our sample. Calibrated on VMP red giants in globular clusters with non-local thermodynamic equilibrium (NLTE) effects considered, this method is well-suited for red giants. We also test its performance on other stellar types, and find that it yields reliable metallicities for MSTO stars, though it is not applicable to dwarf or blue giant stars. Applied to LAMOST spectra, this method enables the identification of a large number of candidate VMP MSTO and red giant stars with robust metallicity estimates.

Using this method, we present a catalog of 8,440 VMP candidates selected from LAMOST DR10, along with their estimated metallicities. Section~\ref{sec:2} describes the dataset used in this study. Section~\ref{sec:3} outlines the candidate selection and metallicity-estimation procedures. In Section~\ref{sec:4}, we present the metallicity results and evaluate their reliability through comparisons with high-resolution spectroscopic samples, other spectroscopic surveys, and machine-learning-based estimates. Section~\ref{sec:5} summarizes our findings and discusses the potential of this approach for future studies.

\section{Data \label{sec:2}}

In this study, we primarily utilize spectroscopic data from LAMOST DR10 \citep{LAMOST} and photometric and astrometric data from \textit{Gaia} DR3 \citep{Gaia}. LAMOST currently has the world’s largest (ground-based) stellar spectral database, and its latest public release, DR10, provides over 12 million stellar spectra in the Milky Way. With its wide wavelength coverage (3700--9100\,\AA), LAMOST enables robust determination of stellar parameters and elemental abundances.

To ensure data quality, we selected low-resolution LAMOST spectra with SNR greater than 20 in the $r$-band, which ensures reliable measurements, particularly in the near-infrared. For stars with multiple observations, we retained the spectrum with the highest SNR. These were then cross-matched with \textit{Gaia} DR3 sources to incorporate magnitudes and distances, adopting the geometric distance estimates from \citet{Gaiadist}.

\section{Methods \label{sec:3}}

\begin{figure*}
\centering
\plotone{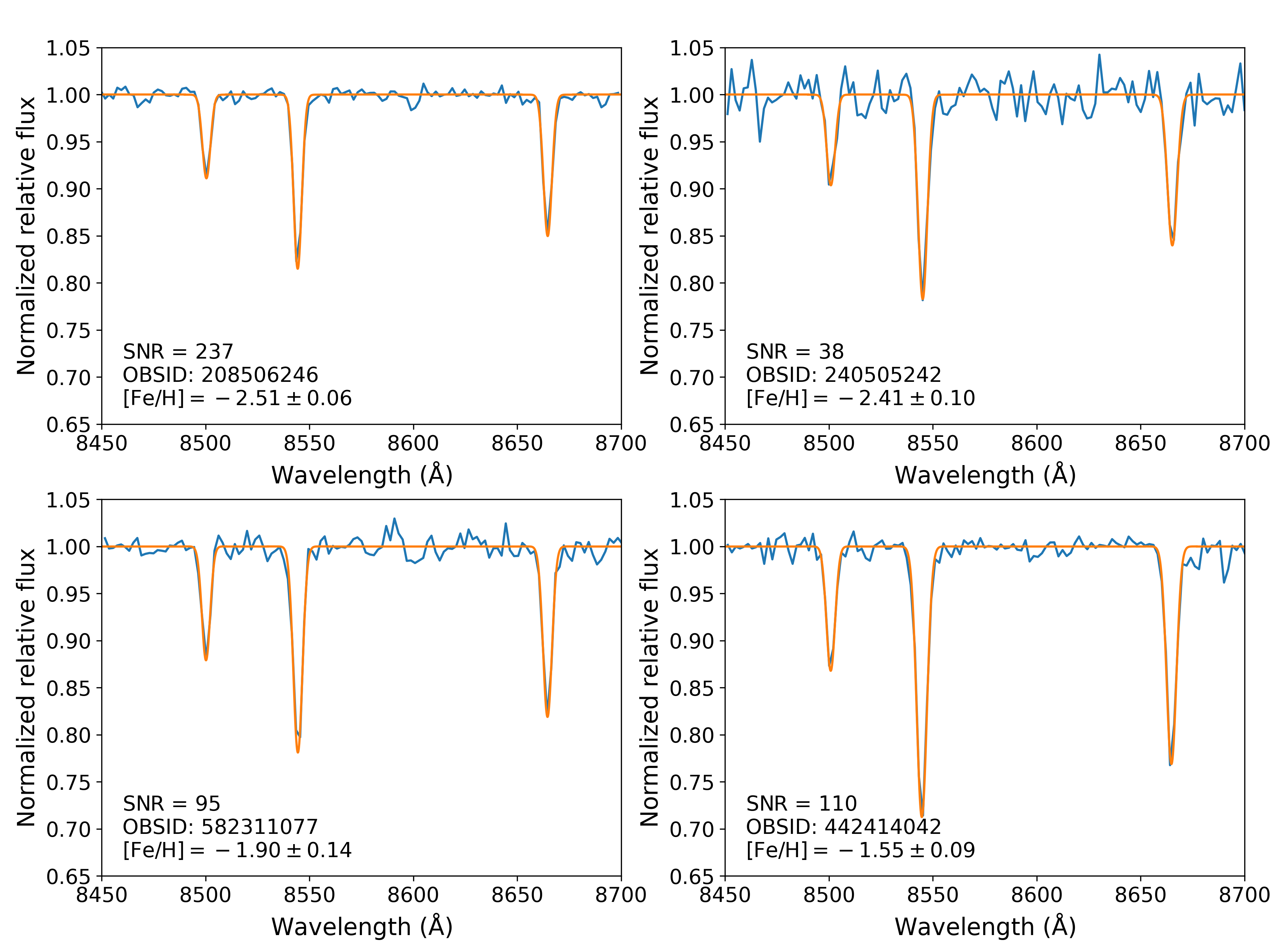}
\caption{Normalized spectra and fitted CaT lines for four example stars in this work. OBSID stands for LAMOST DR10 unique spectrum ID. These examples cover the main range of metallicities and SNRs of our samples. The blue line represents the normalized spectra, and the orange line represents the fitted CaT lines.}
\label{fig:spectra}
\end{figure*}

\begin{figure*}
\centering
\plotone{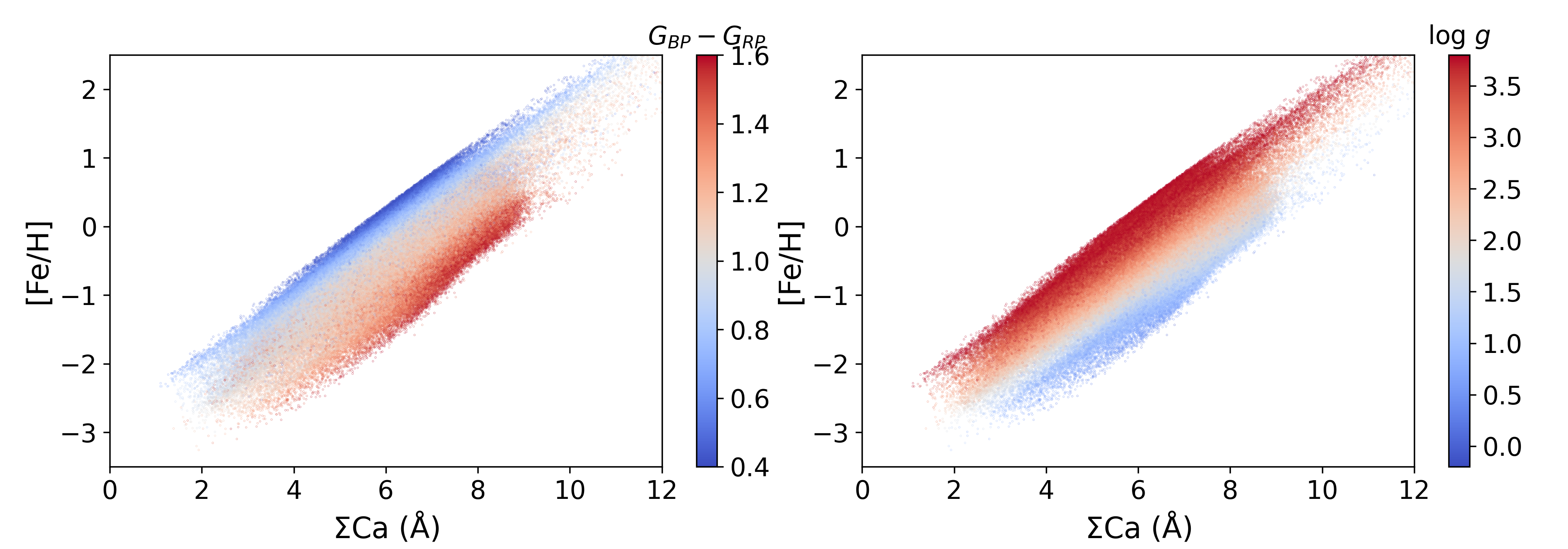}
\caption{The correlation between metallicity and the CaT index, \textit{Gaia BP-RP} color (left), and surface gravity (right). The metallicities are estimated by Method 1, while the surface gravity values are estimated by LASP. This color map shows that metallicity has a strong relationship with the CaT index, \textit{Gaia BP-RP} color, and surface gravity.}
\label{fig:color}
\end{figure*}

\begin{figure}
\centering
\plotone{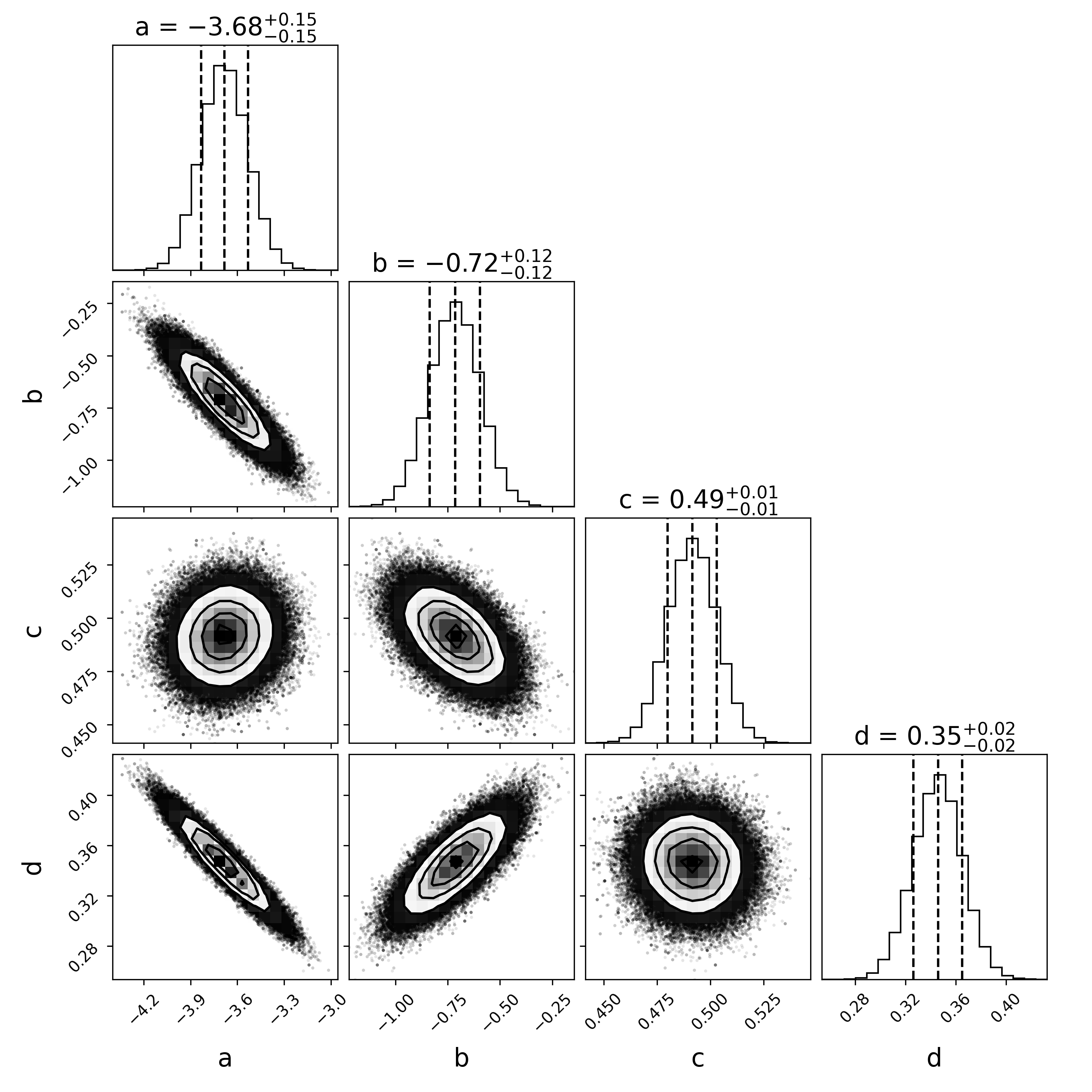}
\caption{The MCMC result for the coefficients in Equation \ref{eq:2}. The values at the top of the columns represent the median offset and one-sigma standard deviation of the coefficients. This figure shows good fitting results, indicating that the coefficients are reliable.}
\label{fig:mcmc}
\end{figure}

\subsection{Method 1: The Absolute Magnitude Method \label{subsec:3.1}}

The absolute magnitude method (hereafter referred to as Method 1) is based on the calibration provided by \citet{Carrera}, which requires CaT EWs and absolute magnitudes as inputs, as expressed in Equation \ref{eq:1}. We found that \citet{Carrera} offered four sets of coefficients for different photometric systems, and the results obtained using these four coefficient sets all showed good consistency. Therefore, we chose the coefficient set using Johnson-Cousins \textit{V} magnitudes as input. Some of the candidates did not have precise \textit{V} magnitudes, so we used the conversion provided by \citet{Gaiamag} to convert \textit{Gaia G} magnitudes to \textit{V} magnitudes. The coefficients are shown in the left column of Table \ref{tab:1}.
\begin{equation}
\mathrm{[Fe/H]} = a + b \times M_V + c \times \Sigma Ca + d \times \Sigma Ca^{-1.5} + e \times \Sigma Ca \times M_V
\label{eq:1}
\end{equation}

\subsubsection{Candidate Selection \label{subsec:3.1.1}}

The calibration from \citet{Carrera} is intended for red giant stars, but we also tested its performance on other stellar types and compared the results with metal-poor samples from high-resolution spectroscopy. We find that this calibration yields reliable metallicities for MSTO stars with absolute magnitudes brighter than $G \sim 6.5$, though it is not applicable to dwarf or blue giant stars. Thus, we adopt an empirical cut from \citet{Huang1} based on \textit{Gaia G} magnitudes and \textit{Gaia BP-RP} color: $0.4 < G_{BP}-G_{RP} < 1.6$ and $-4.0 < M_G < 6.5$, to select MSTO and red giant stars. The cut is shown as the black rectangle on the color-magnitude diagram in Figure \ref{fig:HR}. We then cross-matched the selected MSTO and red giant stars with the LAMOST stellar parameter catalog, derived from LAMOST low-resolution spectra using the LAMOST stellar parameter pipeline \citep[LASP;][]{AFGK}. This catalog presents metallicities for stars in LAMOST DR10 with a selection criterion of [Fe/H] $>-$2.5. We further exclude stars with [Fe/H] $>-$1.5, as they are unlikely to be VMP stars. The final sample comprises approximately 200,000 MSTO and red giant stars with [Fe/H] $<-$1.5.

\subsubsection{Spectral Metallicity Estimate Pipeline \label{subsec:3.1.2}}

First, we shifted all LAMOST spectra to the rest frame using radial velocities derived from LASP, then manually inspected for anomalous spectra and corrected their radial velocities using an empirical template spectrum. Next, we trimmed the spectra to the range 8450–8700\AA\ to focus on the CaT lines, and masked the major absorption features following the approach in \citet{Gaiafeh}. We then fitted a cubic polynomial model to the unmasked regions of the spectra, using the {\tt\string curvefit} function from the {\tt\string SciPy} module in {\tt\string Python}.

After continuum normalization, we modeled the CaT lines separately using Gaussian profiles. We performed the fitting using the {\tt\string LevMarLSQFitter} function from the {\tt\string Astropy} module in {\tt\string Python}. Constraints were applied to each line to ensure that the fitting parameters were reasonable. We selected 1,164 VMP stars with high SNR to validate this fitting method. The left panel of Figure \ref{fig:ew} shows that the EWs derived using Gaussian fitting agree well with direct integration results. In cases where the SNR was too low for Gaussian fitting, we directly integrated the spectra within $\pm$10\AA, $\pm$15\AA, and $\pm$12.5\AA\ around the line cores of the CaT lines to estimate their EWs.

We defined the CaT index ($\Sigma$Ca) as the unweighted sum of the three lines $W_{8500}+W_{8544}+W_{8664}$. When the weakest CaT line $W_{8500}$ was buried in the noise, we defined the CaT index as 1.24 times the sum of the other two lines, $1.24 \times (W_{8544}+W_{8664})$. We classified the CaT indices using two distinct flags: the CaT indices measured with all three CaT lines were assigned {\tt\string flag=0}, while those measured with the two redder lines received {\tt\string flag=1}. The right panel of Figure \ref{fig:ew} demonstrates that these two definitions are consistent for VMP stars. The CaT indices were then converted to metallicities using Equation \ref{eq:1}.

\begin{figure*}
\centering
\plotone{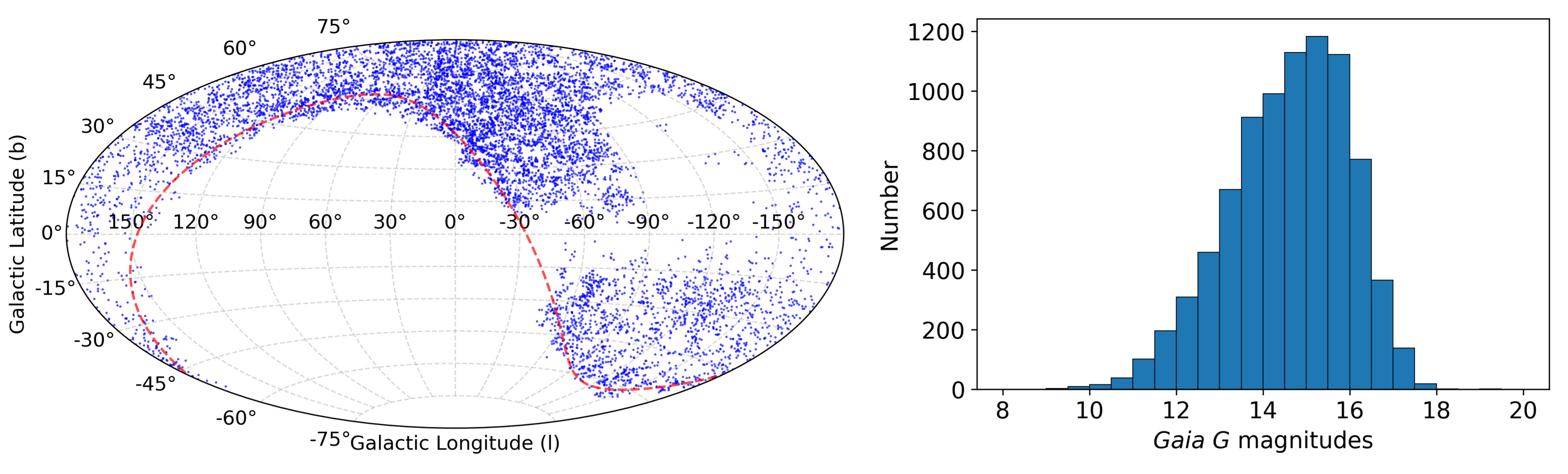}
\caption{Left: Distribution of the VMP sample in Galactic coordinates. The red-dashed line represents the celestial equator. This VMP sample covers a large area of the sky in the Northern Hemisphere. Right:  Distribution of the magnitudes of the VMP sample. More than 7,000 stars in this VMP catalog are brighter than $G \sim 16$.}
\label{fig:sky}
\end{figure*}

\subsection{Method 2: The Color and Surface Gravity Method \label{subsec:3.2}}

Since Method 1 requires absolute magnitudes as input, it heavily relies on accurate distances. However, the geometric distances from \citet{Gaiadist} have significant systematic errors for stars beyond 6 kpc. In order to discover more distant VMP stars, we adopted a new method (hereafter referred to as Method 2), using the \textit{Gaia BP-RP} color and surface gravity log\,$g$ instead of absolute magnitude to measure metallicity. In Figure \ref{fig:color}, we adopted a sample with surface gravity values estimated by LASP and metallicities estimated by Method 1 to show that metallicity has a good relationship with the CaT index, color, and surface gravity.

We refined the calibration from \citet{Carrera} by replacing the three terms involving magnitudes with one term for color and another term for surface gravity, as shown in Equation \ref{eq:2}. The coefficients in this equation are shown in the right column of Table \ref{tab:1}.
\begin{equation}
\mathrm{[Fe/H]} = a + b \times (G_{BP}-G_{RP}) + c \times \Sigma Ca + d \times \mathrm{log\,}g
\label{eq:2}
\end{equation}

The coefficients in Equation \ref{eq:2} were determined using a Markov Chain Monte Carlo (MCMC) procedure based on a metal-poor sample with metallicities estimated by Method 1 and surface gravity values estimated by LASP. This sample allows us to minimize the systematic errors between the two methods. Figure \ref{fig:mcmc} exhibits good MCMC fitting results, indicating that the coefficients are reliable.

Since we no longer have accurate distances for distant stars, we replaced the absolute magnitude cut with a surface gravity cut of log\,$g <$ 3.8, while retaining the color cut of $0.4 < G_{BP}-G_{RP} < 1.6$ to select MSTO and red giant stars. We use the two-dimensional dust map from \citet{SFD} for de-reddening the \textit{Gaia BP-RP} color. This dust map is generally consistent with the dust map from \citet{Green19}. We removed stars with [Fe/H] $> -$1.5 from the LAMOST stellar parameter catalog, as they are unlikely to be VMP stars. The final sample comprises approximately 20,000 MSTO and red giant stars with [Fe/H] $< -$1.5, other than those in the the sample from Method 1. We used the same procedure as Method 1 for radial velocity correction, continuum normalization, spectral fitting, and CaT index measurements.

\section{Results and Validation \label{sec:4}}

\begin{figure*}
\centering
\plotone{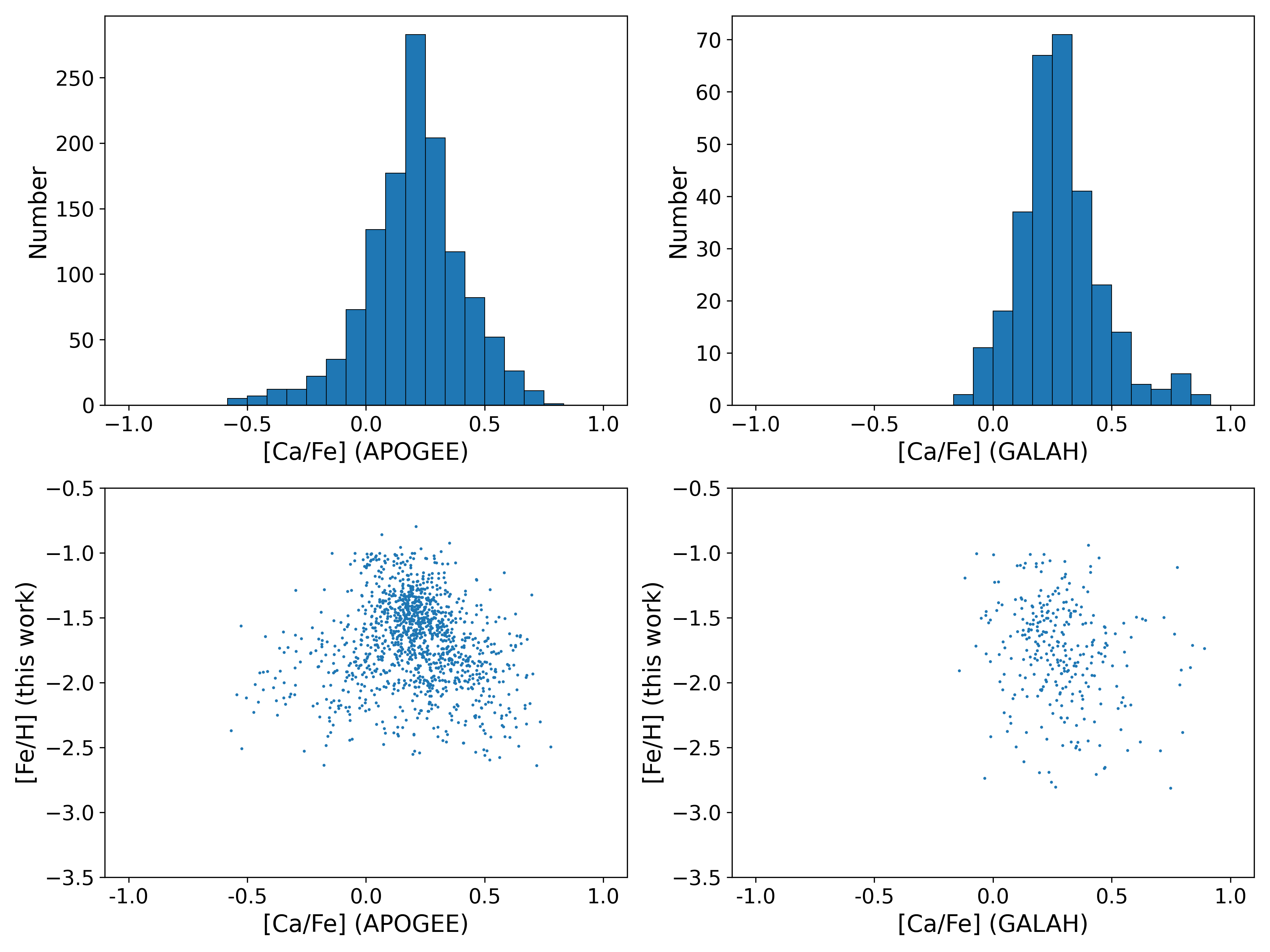}
\caption{Top row: The [Ca/Fe] distribution of metal-poor stars in this work using measurements from APOGEE (left) and GALAH (right). Bottom row: The distribution of [Fe/H] measured in this work versus the [Ca/Fe] measured by APOGEE (left) and GALAH (right).}
\label{fig:ca}
\end{figure*}

We have used the methods described above to estimate the metallicities of $\sim$220,000 MSTO and red giant stars in LAMOST DR10, including 8,440 VMP stars. Among these stars, 4,500 metallicity estimates were obtained using Method 1, and 3,940 estimates were obtained using Method 2. Figure \ref{fig:sky} shows that this catalog covers a large area of the sky in the Northern Hemisphere, with metallicities ranging from [Fe/H] $=-$2.0 to $-$4.0, including 25 extremely metal-poor (EMP; [Fe/H] $\leq -$3.0) stars. More than 7,000 stars are brighter than $G \sim 16$, making them excellent targets for high-resolution spectroscopic follow-up with 4–10 meter-class telescopes.

We have validated our metallicity estimates by comparing them with the \textit{Gaia} RVS metallicity catalog \citep{Gaiafeh}, spectroscopic surveys such as APOGEE DR17 \citep{APOGEE} and GALAH DR3 \citep{GALAH}, high-resolution spectroscopic databases such as PASTEL \citep{PASTEL} and SAGA \citep{SAGA}, metal-poor samples from previous works such as \citet{Subaru}, and metallicity catalogs derived using the DD-PAYNE method \citep{DDPAYNE} and a neural network method \citep{VAC}. We only included samples with [Fe/H] $< -$1.5 for comparison because the calibration used in this work is only suitable for VMP stars, and our metallicity estimates have relatively large systematic errors when compared to literature values for samples with [Fe/H] $> -$1.5.

\subsection{Comparison with Metal-poor Samples from \textit{Gaia} RVS Spectroscopy \label{subsec:4.1}}

Figure \ref{fig:cross0} shows the validation of the metallicities estimated by Method 1. The metal-poor samples from \textit{Gaia} RVS spectroscopy \citep{Gaiafeh} used the same calibration as in this work, providing an opportunity to validate our spectral analysis pipeline without introducing systematic errors from the calibration. In addition to comparing metallicities, we also compared the CaT EWs measured in both works. 

From inspection of the top row of Figure \ref{fig:cross0}, one can see that the metallicity estimates exhibit a negligible median offset of less than +0.01 dex, and a standard deviation of 0.13 dex, while the CaT indices show a moderate median offset of $-$0.11 dex, and a standard deviation of 0.25 dex. This demonstrates the robustness of our spectral analysis pipeline.

Figure \ref{fig:cross1} shows the validation of the metallicities estimated by Method 2. In the top-left panel we compared the metallicities estimated by Method 1 and Method 2, while in the top-right panel we compared the metallicities estimated by Method 2 and \textit{Gaia} RVS spectroscopy \citep{Gaiafeh}. Both comparisons exhibit very small median offsets of $+$0.01 and $-$0.01 dex, and moderate standard deviations of 0.15 and 0.18 dex, indicating that the calibration based on color and surface gravity is a good alternative to the calibration provided by \citet{Carrera}.

\subsection{Comparison with Metal-poor samples from Spectroscopic Surveys \label{subsec:4.2}}

Next, we validated our metallicities by comparing them with metal-poor samples from spectroscopic surveys such as APOGEE DR17 \citep{APOGEE} and GALAH DR3 \citep{GALAH}. We removed stars with {\tt\string FE$\_$H$\_$FLAG$\neq$0} from the APOGEE sample and stars with {\tt\string flag$\_$fe$\_$h$\neq$0} from the GALAH sample.

From inspection of the second row of Figure \ref{fig:cross0}, one can see that the metallicities estimated by Method 1 exhibit a tiny median offset of less than 0.01 dex, and a moderate standard deviation of 0.13 dex when compared to the APOGEE sample, while showing a slightly larger, but still very small, median offset of $-$0.08 dex, and a moderate standard deviation of 0.18 dex when compared to the GALAH sample. This indicates that our metallicity estimates are reliable for VMP stars.

The second row of Figure \ref{fig:cross1} shows that the metallicities estimated by Method 2 exhibit a tiny median offset of less than +0.01 dex, and a moderate standard deviation of 0.16 dex when compared to the APOGEE sample, while showing a larger median offset of $-$0.14 dex, and a moderate standard deviation of 0.18 dex when compared to the GALAH sample. Generally, the comparison of metallicities estimated by Method 1 and Method 2 have similar trends, probably because the coefficients of the calibration used in Method 2 were derived with a sample from Method 1, and thus Method 2 inherited the systematic offsets from Method 1.

We also validated the calcium abundance for the metal-poor stars in this work using [Ca/Fe] from APOGEE and GALAH. The top row of Figure \ref{fig:ca} demonstrates that the [Ca/Fe] of most metal-poor stars are between 0.0 and $+$0.4, while the bottom row of Figure \ref{fig:ca} shows no significant correlation between the [Ca/Fe] and [Fe/H] of these stars. Given this relatively narrow spread and lack of strong correlation, we do not apply any explicit correction for calcium abundance in our [Fe/H] calibration based on CaT lines. However, we adopt a conservative empirical uncertainty of 0.2 dex to account for possible variations due to [Ca/Fe].

\subsection{Comparison with Metal-poor Samples from High-resolution Spectroscopy \label{subsec:4.3}}

To validate our metallicities in the EMP regime, we compared them with metal-poor samples from high-resolution spectroscopic databases such as PASTEL \citep{PASTEL} and SAGA \citep{SAGA}, as well as results from previous works based on high-resolution spectroscopy, such as \citet{Subaru}.

Inspection of the third row of Figure \ref{fig:cross0} shows that the metallicity estimates in this work exhibit moderate median offsets of $+$0.21 and $+$0.20 dex, and standard deviations of 0.28 and 0.25 dex when compared to high-resolution results. The metallicities estimated by Method 1 are higher in the EMP regime than in the VMP regime, because the calibration used in Method 1 saturate in the EMP regime \citep{Carrera}.

The third row of Figure \ref{fig:cross1} also exhibits moderate median offsets of $+$0.11 and $+$0.25 dex, and standard deviations of 0.25 and 0.30 dex. This comparison lacks EMP stars because we used surface gravity values estimated by LASP, which did not assign stellar parameters to EMP stars. However, we will be able to identify EMP stars using this method once we will have access to additional sources of surface gravity data in the future.

\subsection{Comparison with Metallicities Estimated by DD-Payne and Neural Network Methods \label{subsec:4.4}}

\begin{figure*}
\centering
\includegraphics[scale=0.45]{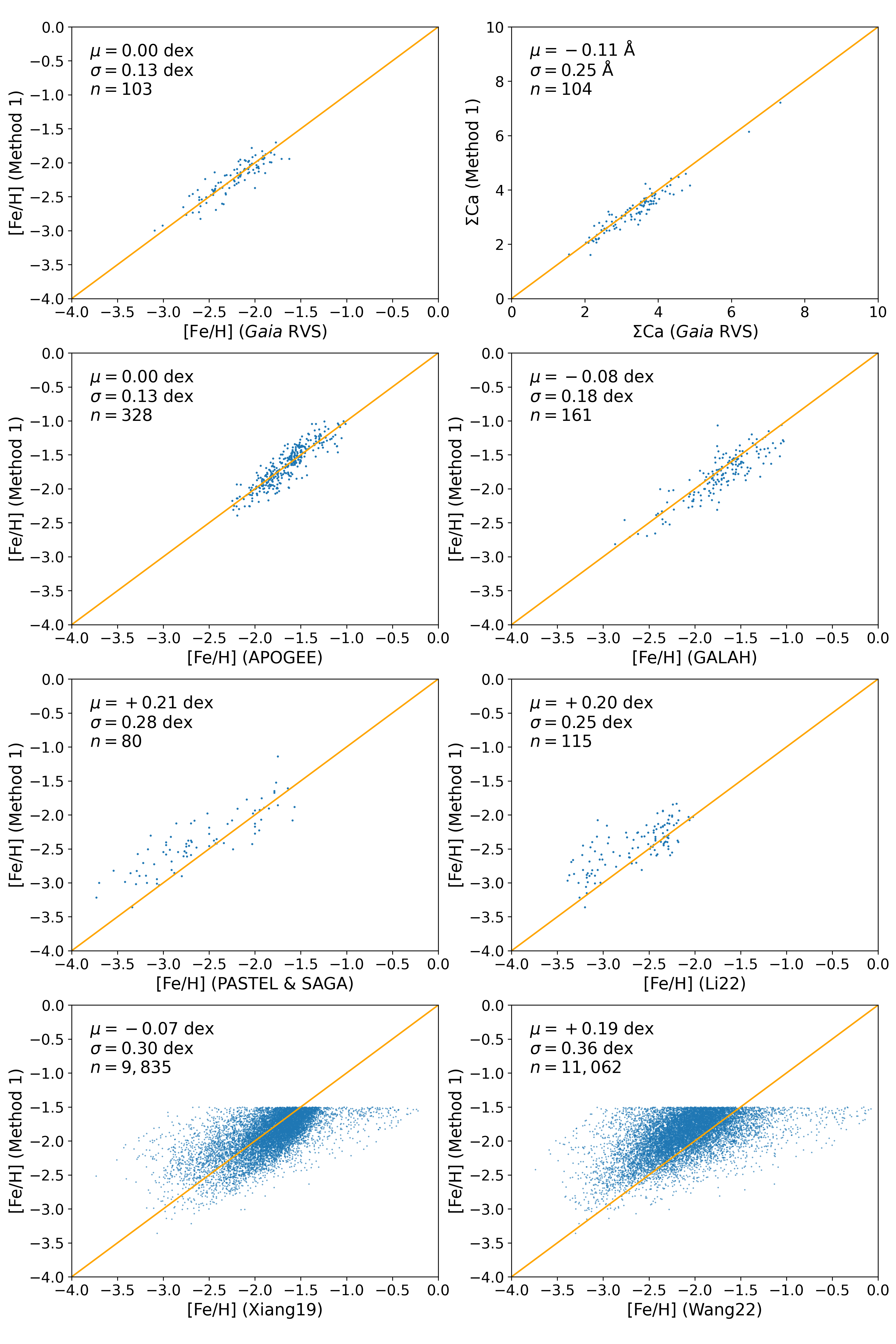}
\caption{Top row: Comparison of metallicities (left) and CaT EWs (right) estimated by Method 1 with results from \textit{Gaia} RVS spectroscopy. Second row: Comparison with spectroscopic surveys such as APOGEE DR17 (left) and GALAH DR3 (right). Third row: Comparison with metal-poor samples from high-resolution spectroscopy such as PASTEL \& SAGA (left) and \citet{Subaru} (right). Bottom row: Comparison with metallicity catalogs from \citet{DDPAYNE} (left) and \citet{VAC} (right). The number of matches (n), median offset ($\mu$), and standard deviation ($\sigma$) of [Fe/H] and $\Sigma$Ca are shown at the top of each panel. The orange lines represent the 1:1 line.}
\label{fig:cross0}
\end{figure*}

The bottom row of Figure \ref{fig:cross0} shows comparisons with metal-poor samples from \citet{DDPAYNE} using the data-driven Payne (DD-Payne) method, and metal-poor samples from \citet{VAC} using a neural network method. The DD-Payne model was trained with samples from GALAH DR2 \citep{GALAH2} and APOGEE DR14 \citep{APOGEE14}, while the neural network model was trained with samples from PASTEL \citep{PASTEL}. 

\begin{figure*}
\centering
\includegraphics[scale=0.45]{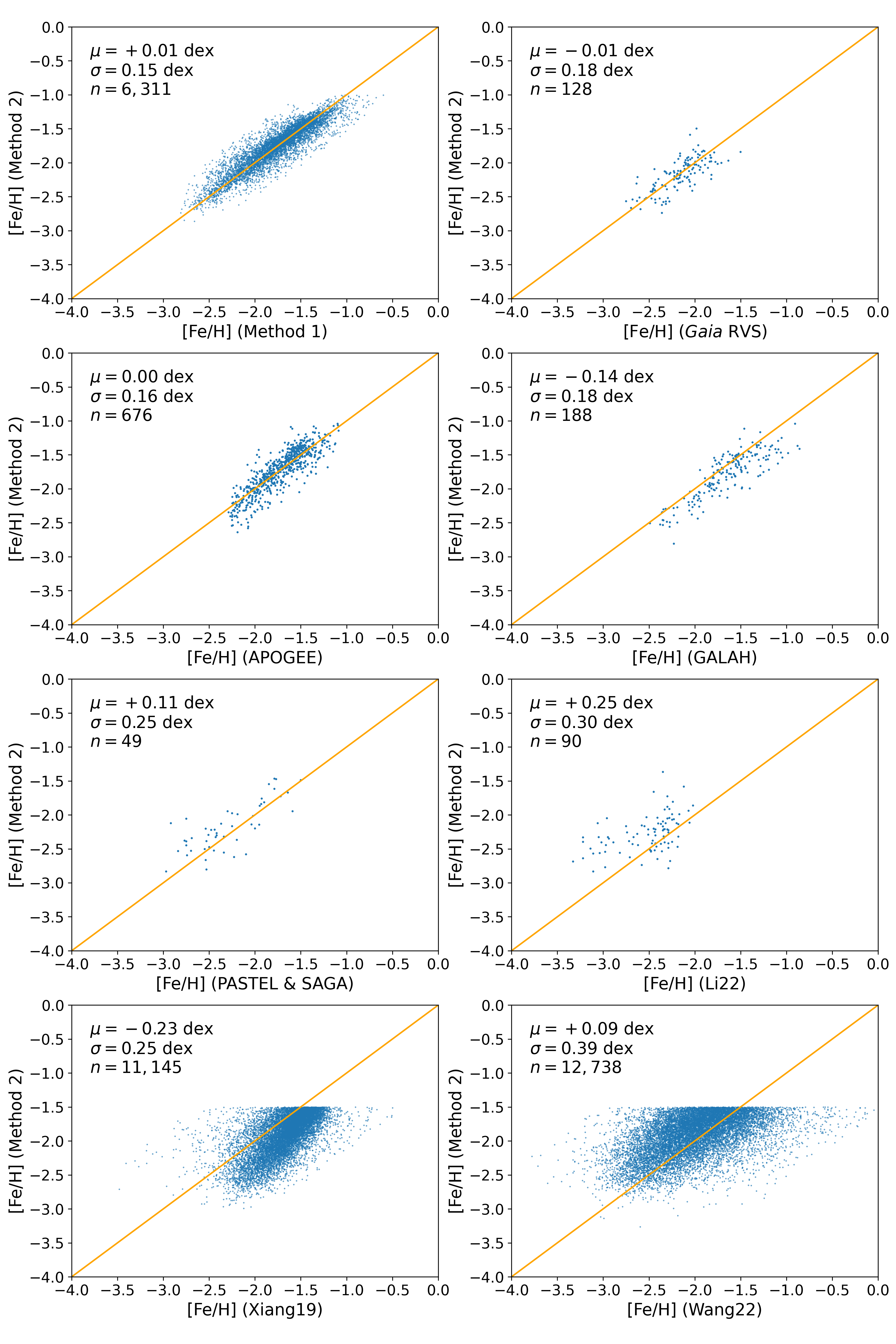}
\caption{Top row: Comparison between metallicities estimated by Method 1 and Method 2 (left) and comparison between metal-poor samples from Method 2 and \textit{Gaia} RVS spectroscopy (right). Second row: Comparison with spectroscopic surveys such as APOGEE DR17 (left) and GALAH DR3 (right). Third row: Comparison with metal-poor samples from high-resolution spectroscopy such as PASTEL and SAGA (left) and \citet{Subaru} (right). Bottom row: Comparison with metallicity catalogs from \citet{DDPAYNE} (left) and \citet{VAC} (right). The number of matches (n), median offset ($\mu$), and standard deviation ($\sigma$) of [Fe/H] are shown at the top of each panel. The orange lines represent the 1:1 line.}
\label{fig:cross1}
\end{figure*}

The bottom row of Figure \ref{fig:cross0} exhibits a very small median offset of $-$0.07 dex, and a standard deviation of 0.30 dex compared to \citet{DDPAYNE}, and a moderate median offset of +0.19 dex, and a standard deviation of 0.36 dex compared to \citet{VAC}. The bottom row of Figure \ref{fig:cross1} exhibits similar results, with a larger median offset of $-$0.23 dex, and a standard deviation of 0.25 dex compared to \citet{DDPAYNE}, and a smaller median offset of +0.09 dex, and a standard deviation of 0.39 dex compared to \citet{VAC}.

The bottom row of this figure shows greater dispersions than the upper rows, probably due to the combined errors from both our sample and the sample from \citet{DDPAYNE} and \citet{VAC}. The systematic offset in the bottom-left panel occurs because the DD-Payne model saturates at [Fe/H] $< -$2.5, while the offset in the bottom-right panel arises from the difference between low-resolution spectra from LAMOST and the high-resolution spectra from PASTEL. Overall, the metallicities estimated in this work agree with the results from \citet{DDPAYNE} and \citet{VAC}.

\begin{table*}
\caption{Description of the Columns in the Metallicity Catalog in this Work}
\begin{tabular}{lll}
\hline
\textbf{Column} & \textbf{Description} & \textbf{Unit}\\
\hline
obsid & LAMOST DR10 unique spectrum ID & \dots\\
gaiaid & \textit{Gaia} DR3 Source ID & \dots\\
ra & \textit{Gaia} DR3 right ascension in ICRS (J2016) format & degrees\\
dec & \textit{Gaia} DR3 declination in ICRS (J2016) format & degrees\\
snr & Signal-to-noise ratio at r-band of the LAMOST spectrum & \dots\\
g\_mag & \textit{Gaia G} magnitude & mag\\
g\_err & Uncertainty of the \textit{Gaia G} magnitude & mag\\
bp\_mag & \textit{Gaia} $G_{BP}$ magnitude & mag\\
bp\_err & Uncertainty of the \textit{Gaia} $G_{BP}$ magnitude & mag\\
rp\_mag & \textit{Gaia} $G_{RP}$ magnitude & mag\\
rp\_err & Uncertainty of the \textit{Gaia} $G_{RP}$ magnitude & mag\\
teff\_LASP & Effective temperature estimated by \citet{AFGK} & K\\
teff\_err\_LASP & Uncertainty of the effective temperature estimated by \citet{AFGK} & K\\
logg\_LASP & Surface gravity estimated by \citet{AFGK} & cgs\\
logg\_err\_LASP & Uncertainty of the surface gravity estimated by \citet{AFGK} & cgs\\
feh\_LASP & Metallicity estimated by \citet{AFGK} & \dots\\
feh\_err\_LASP & Uncertainty of the metallicity estimated by \citet{AFGK} & dex\\
pm & \textit{Gaia} DR3 proper motion & mas yr$^{-1}$\\
pmra & \textit{Gaia} DR3 proper motion in right ascension & mas yr$^{-1}$\\
pmra\_err & Uncertainty of the \textit{Gaia} DR3 proper motion in right ascension & mas yr$^{-1}$\\
pmdec & \textit{Gaia} DR3 proper motion in declination & mas yr$^{-1}$\\
pmdec\_err & Uncertainty of the \textit{Gaia} DR3 proper motion in declination & mas yr$^{-1}$\\
rv & Radial velocity offset from the rest frame & km s$^{-1}$\\
rv\_err & Uncertainty of the radial velocity offset from the rest frame & km s$^{-1}$\\
rgeo & Geometric distance estimated by \citet{Gaiadist} & pc\\
rgeo\_err & Uncertainty of the geometric distance estimated by \citet{Gaiadist} & pc\\
ebv & Interstellar reddening & mag\\
g\_corr & De-reddened \textit{Gaia G} magnitude & mag\\
bp\_corr & De-reddened \textit{Gaia} $G_{BP}$ magnitude & mag\\
rp\_corr & De-reddened \textit{Gaia} $G_{RP}$ magnitude & mag\\
ew1 & Equivalent width of the first calcium triplet line at 8500\AA & \AA\\
err1 & Uncertainty of the equivalent width of the first calcium triplet line at 8500\AA & \AA\\
chi1 & The $\chi^2$ of the spectra fitting of the first calcium triplet line at 8500\AA & \dots\\
ew2 & Equivalent width of the second calcium triplet line at 8544\AA & \AA\\
err2 & Uncertainty of the equivalent width of the second calcium triplet line at 8544\AA & \AA\\
chi2 & The $\chi^2$ of the spectra fitting of the second calcium triplet line at 8544\AA & \dots\\
ew3 & Equivalent width of the third calcium triplet line at 8664\AA & \AA\\
err3 & Uncertainty of the equivalent width of the third calcium triplet line at 8664\AA & \AA\\
chi3 & The $\chi^2$ of the spectra fitting of the third calcium triplet line at 8664\AA & \dots\\
err & The average noise of the continuum in the spectrum & \AA\\
flag & Quality flag of the equivalent width of the first calcium triplet line at 8500\AA & \dots\\
feh & Spectroscopic metallicity derived in this work using LAMOST spectrum & \dots\\
feh\_err & Uncertainty of the spectroscopic metallicity derived in this work & dex\\
method & Method used in deriving spectroscopic metallicity & \dots\\
\hline
\end{tabular}
\label{tab:2}
\end{table*}

\section{Summary \label{sec:5}}
    
The search for very metal-poor (VMP) stars has long been a challenging yet central pursuit in Galactic Archaeology. In the past decade, data mining of photometric and spectroscopic surveys has led to numerous successful discoveries of VMP stars in the Galaxy. LAMOST DR10 has released over 12 million spectra for stars in the Milky Way, providing a great opportunity for discovering VMP stars. Unlike previous studies, we focus on the red arm of LAMOST spectra because its high SNR enables the analysis of a larger sample size and more accurate metallicity estimates. We derived metallicities using the equivalent widths (EWs) of the Calcium Triplet (CaT) lines in the near-infrared band of LAMOST low-resolution spectra, applying an empirical calibration from \citet{Carrera}. Additionally, we refined this calibration by replacing absolute magnitude with \textit{Gaia BP-RP} color and LAMOST surface gravity, facilitating the discovery of distant VMP stars without the need for accurate distance measurements.

In this work, we present a metallicity catalog of 8,440 VMP MSTO and red giant stars based on low-resolution spectra from LAMOST DR10. The primary parameters of the catalog are listed in Table \ref{tab:2}. This catalog covers a large area of sky in the Northern Hemisphere, with metallicities ranging from [Fe/H] $=-$2.0 to $-$4.0. In addition to metallicities, we also provide distances, proper motions, and radial velocities from which complete six-dimensional dynamic data can be derived for the sample stars. About 4000 stars are located farther than 6 kpc, and more than 7,000 stars are brighter than $G \sim 16$. These bright VMP stars are excellent targets for high-resolution spectroscopic follow-up with 4–10 meter-class telescopes. Validity checks through comparisons with metal-poor samples from spectroscopic surveys and high-resolution spectroscopy demonstrate that the metallicity estimates in this work are robust and reliable, with an average median offset of $\sim$0.1 dex and an average standard deviation of $\sim$0.2 dex.

The ongoing data releases from the \textit{Gaia} project will provide accurate parallaxes and proper motions for billions of stars in the Milky Way, which are expected to significantly improve the accuracy of stellar parameters, especially surface gravity. Moreover, future data releases from spectroscopic surveys such as LAMOST DR12 and DESI DR1 will provide spectra for millions of stars in the Galaxy. The combination of future surveys will result in an unprecedented multi-parameter catalog of Galactic stars, offering an enormous opportunity to explore the nature of the early Milky Way.

\begin{acknowledgments}
Y.H. acknowledges support by the National Natural Science Foundation of China grants No. 12422303 and the National Key R\&D Program of China No. 2023YFA1608303.
H.W.Z. acknowledges support by the National Key R\&D Program of China No. 2024YFA1611903.
T.C.B. acknowledges partial support for this work from grant PHY 14-30152; Physics Frontier Center/JINA Center for the Evolution of the Elements (JINA-CEE), and OISE-1927130: The International Research Network for Nuclear Astrophysics (IReNA), awarded by the US National Science Foundation.
Guoshoujing Telescope (the Large sky Area Multi-Object Fiber Spectroscopic Telescope, LAMOST) is a National Major Scientific Project built by the Chinese Academy of Sciences. Funding for the project has been provided by the National Development and Reform Commission. LAMOST is operated and managed by the National Astronomical Observatories, Chinese Academy of Sciences.

\textit{Software:} NumPy\citep{NumPy}, Matplotlib\citep{Matplotlib}, SciPy\citep{SciPy}, Astropy\citep{Astropy}, TOPCAT\citep{TOPCAT}

\end{acknowledgments}

\bibliography{paper}{}
\bibliographystyle{aasjournal}

\end{document}